**ORIGINAL ARTICLE**

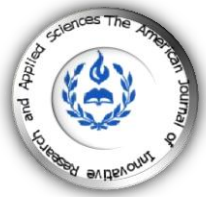

# EFFET DU CHLORURE DE SODIUM (NaCl) SUR LA CROISSANCE DE SIX ESPECES d'*Acacia*

# EFFECT OF SODIUM CHLORIDE (NaCl) ON THE GROWTH OF SIX *Acacia* SPECIES

|Khalil Chérifi [1]\* |Abdelmjid Anagri [1] | El Houssine Boufous [2] | and | Abelhamid El Mousadik [1] |

[1.] Laboratory of Biotechnology and Valorization of Natural Resources | Faculty of sciences | Ibn Zohr University | P.O. Box 8106 | 8000 Agadir | Morocco |
[2.] Department of Biochemistry and Microbiology | Laval University | Quebec City (Quebec) | Canada |

**RESUME**

**Introduction :** Au cours de ces dernières décennies on assiste à une diminution progressive des superficies cultivables dans les régions arides et semi-arides à cause de l'accumulation des sels liée à la rareté des précipitations, au mauvais drainage, à la sècheresse prolongées et à l'absorption de l'eau par les plantes. **Contexte :** Devant l'ampleur de ce problème, il s'avère donc nécessaire de proposer des programmes d'évaluation et de conservation des espèces menacées d'extinction. Le repérage d'espèces plus adaptées et la sélection des variétés tolérantes à la salinité resteraient la voie économique la plus efficace pour l'exploitation des terrains affectés. **Objectifs :** L'objectif de cette étude est de déterminer la capacité de tolérance à la salinité au cours du développement végétatif chez six espèces appartenant au genre *Acacia*. Ceci dans le but d'élaborer une classification des seuils de tolérance au stress salin, critère important dans le choix des espèces à retenir dans un programme de mise en valeur des zones affectées par la salinité. **Méthodes :** L'effet du stress salin a été abordé sur un certain nombre de caractères agro-morphologiques en conditions contrôlées. Les concentrations de NaCl appliquées, en plus du témoin, sont : 100 mM, 200 mM, 300 mM et 400 mM. **Résultats :** Les résultats ont montré une variabilité non négligeable dans le comportement des plantes des différentes espèces en fonction du stress salin. Chez les six espèces d'*Acacia*, le sel a exercé un effet dépressif sur tous les paramètres de croissance étudiés. Toutefois, le taux de réduction diffère selon l'intensité de stress salin et le degré de sensibilité ou de tolérance de l'espèce. La croissance en hauteur, le nombre de feuilles et la biomasse sèche totale sont vraisemblablement les paramètres les plus affectés. Cependant, il est important de signaler que toutes les espèces d'*Acacia* considérées dans ce travail ont survécu, même à 400 mM de NaCl, et ont présenté différents degrés de tolérance à la salinité. Dans cette situation, les espèces *A. horrid*a et *A. raddiana* s'avèrent globalement les plus performantes au stade végétatif. **Conclusions :** La variabilité génétique, dévoilée par ces espèces dans les différentes conditions de stress salin, permettrait un choix d'écotypes pouvant entrer dans des schémas de sélection et d'amélioration variétale pour la réhabilitation des parcours dégradés surtout en zones affectées par la salinité.
**Mots-clé**s: *Tolérance à la salinité, Variabilité, Acacia, Amélioration des plantes, Réhabilitation*

**ABSTRACT**

**Background:** Salinity is one of the major abiotic stresses affecting plant production in arid and semi-arid regions. It causes reduction of cultivable area and combined with other factors, presents a serious threat to food stability in these areas. **Context:** In front of this problem, the selection of salt tolerant species and varieties remains the best economic approach for exploitation and rehabilitation of salt-affected regions. **Objective:** The purpose of this study was to assess and compare the seed germination response of six *Acacia* species under different NaCl concentrations in order to explore opportunities for selection and breeding salt tolerant genotypes. **Methods:** The salinity effect was examined by measuring some agro-morphological parameters in controlled growth environment using five treatment levels: 0, 100, 200, 300 and 400 mM of NaCl. **Results:** The analyzed data revealed significant variability in salt response within and between species. All growth parameters were progressively reduced by increased NaCl concentrations. Growth in height, leaf number and total plant dry weight were considered as the most sensitive parameters. However, the growth reduction varied among species in accordance with their tolerance level. It is important to note that all species survived at the highest salinity (400 mM). Whereas *A. horrida* and *A. raddiana* were proved to be often the best tolerant, they recorded the lowest reduction percentage at this stage. **Conclusion:** The genetic variability found in the studied species at seedling stage may be used to select genotypes particularly suitable for rehabilitation and exploitation of lands affected by salinity.
**Keywords:** *Salt tolerance, Variability, Acacia, Plant breeding, Rehabilitation.*

# 1. INTRODUCTION

La salinisation est un processus important de dégradation des sols. Elle constitue un facteur limitant à la croissance et au développement des plantes. Les conséquences de ce phénomène qui ne cesse de prendre de l'ampleur, se manifestent par la toxicité directe due à l'accumulation excessive des ions ($Na^+$ et $Cl^-$) dans les tissus des organes, et à un déséquilibre nutritionnel imputable essentiellement à des compétitions entre les éléments minéraux, tel que le sodium avec le potassium et le calcium, le chlorure avec le nitrate, le phosphate et le sulfate [1, 2]. En conséquence, les glycophytes les plus tolérantes seront celles qui, tout en utilisant le $Na^+$ comme osmoticum, conserveront une forte sélectivité vis à vis du $K^+$ [3]. La stratégie utilisée par les végétaux pour éviter les problèmes d'excès d'ions tout en réalisant leur équilibre osmotique est la compartimentation cellulaire, qui se traduit par une accumulation préférentielle du $Na^+$ dans la vacuole [4]. Cependant, chez les glycophytes tolérantes, on discerne également une compartimentation à l'échelle de la plante, surtout dans les organes jeunes où la teneur en Na+ reste faible [5, 6].

La variabilité pour la tolérance à la salinité a été étudiée chez beaucoup d'espèces [5, 7-11]. Pour des raisons pratiques, de nombreuses explorations de cette variabilité ont été abordées au stade végétatif sur des plantes très jeunes. Cette approche est justifiée par le fait que la réponse des plantules est parfois fortement prédictive de celle des plantes adultes [12-14]. Dans certains cas, l'écart entre la tolérance au stade plantule et celle au stade adulte peut justifier les différences entre les mécanismes impliqués d'un stade de développement à un autre [15-19].
Plusieurs recherches ont montré que la croissance en hauteur [20, 21], la production de biomasse des tiges et des racines [22, 23] est négativement affectée par l'augmentation de la salinité.

Face à la salinisation des sols qui constitue l'un des facteurs abiotiques majeurs réduisant le rendement agricole, l'introduction d'espèces végétales tolérantes à la salinité est une stratégie alternative recommandée pour valoriser les sols touchés par ce phénomène. Cette approche, permettraient d'améliorer le couvert végétal et résoudre les problèmes de régénération de certaines espèces forestières en zones arides et semi-arides, particulièrement celles appartenant au genre *Acacia* qui représentent certainement une richesse écologique menacée en Afrique du Nord [24].

*Acacia,* appartenant à la famille des Fabacées, est identifié comme étant un genre cosmopolite, varié et riche. Ces espèces à usage multiple peuvent coloniser des sols pauvres grâce à leur capacité de fixer l'azote atmosphérique par leur association symbiotique avec le Rhizobium des nodosités racinaires. Au Maroc les écosystèmes à base d'*Acacia* représentent un enjeu stratégique pour les régions semi-arides, arides et sahariennes du pays aussi bien sur le plan écologique que socio-économique. Ces écosystèmes rares et originaux, peuvent constituer une protection naturel contre la désertification.et fournir un intérêt multifonctionnel et multi-usager, tels que leurs utilisations dans la production de bois d'énergie et de service, de gomme arabique, de substances pharmaceutiques, de fourrage, de produits mellifères, ainsi que leur utilisation en reboisement et en foresterie urbaine [25].
La maîtrise des exigences de croissance des plantules est une étape importante dans le succès des opérations de reboisement de ces espèces. Ce stade, très important pour le développement des plantes, est surtout contrôlé par des facteurs génétiques et environnementales, en particulier la salinité [26]. Malheureusement, au Maroc, peu de travaux de recherche sur le degré de tolérance à la salinité chez les *Acacia* ont été effectués. Notre présente étude s'inscrit dans le cadre d'une évaluation de la variabilité des réponses au stade plantule de six espèces d'*Acacia* soumises à des doses croissantes de NaCl. L'effet du stress salin a été abordé sur un certain nombre de caractères agro-morphologiques en conditions contrôlées. Ceci dans le but d'identifier le matériel végétal le plus performant pour des programmes de restauration et de valorisation de la productivité végétale, particulièrement pour la réhabilitation des parcours dégradés, c'est le cas d'*Acacia gummifera* et *Acacia raddiana*, ainsi que pour des projets de reboisements en milieux affectés par la salinité, dans le cas des espèces exotiques, comprenant *Acacia eburnea*, *Acacia cyanophylla*, *Acacia cyclops* et *Acacia horrida*.

## 2. MATERIELS ET METHODES

### 2.1. Matériel végétal

L'analyse de la diversité de la tolérance à la salinité a porté sur deux espèces autochtones, représentées par *Acacia gummifera* et *Acacia raddiana* et quatre espèces exotiques, représentées par *Acacia eburnea*, *Acacia cyanophylla*, *Acacia cyclops* et *Acacia horrida*. La majeure partie des graines des différentes espèces testées ont été collectées sous forme de gousses dans différentes régions du sud-ouest marocain (Tableau 1). Elles nous ont été aimablement fournies par la station régionale des semences forestières de Marrakech et ont été conservées au froid (4°C) jusqu'à l'analyse.

**Tableau 1** : Le tableau montre la localisation géographique des différents échantillons d'espèces étudiées.

| Espèces | Provenance | Région de provenance |
|---|---|---|
| *A. gummifera* | Reserve de faune de Rmila (Marrakech) | Haut Atlas occidental |
| *A. raddiana* | Reserve de faune de Rmila (Marrakech) | Haut Atlas occidental |
| *A. cyclops* | Région d'Essaouira | Souss Nord |
| *A. cyanophylla* | Canal de Rocade (Marrakech) | Haut Atlas occidental |
| *A. horrida* | Commune rurale de Saada (Marrakech) | Haut Atlas occidental |
| *A. eburnea* | Commune rurale de Saada (Marrakech) | Haut Atlas occidental |

## 2.2. Protocole expérimental

L'expérimentation a été réalisée sous abri (serre en plastique) à la Faculté des Sciences d'Agadir. Les graines, choisies au hasard, au sein des échantillons représentent les différentes espèces, sont scarifiées puis mises à germer dans des boîtes de Pétri, tapissées d'un papier filtre imbibé d'eau distillée. Cinq jours après la mise en gémination, les plantules sont transplantées isolément dans des sacs de culture de 2 litres contenant un mélange de sable (1/3), d'argile (1/3) et de tourbe (1/3). Tous les deux jours, ces plantules sont arrosées à l'eau de robinet à raison de 100 ml par pot. Après de 10 jours, elles ont été soumises au stress salin, et les concentrations de NaCl appliquées, en plus du témoin, sont : 100 mM, 200 mM, 300 mM et 400 mM. Pour éviter le choc osmotique, les doses de NaCl ont été augmentées graduellement jusqu'à la concentration finale [27, 28].

## 2.3. Caractères mesurés

Après huit semaines de culture, quatre paramètres ont été mesurés dans différentes conditions de stress salin (Tableau 2). Les caractères retenus se rapportent au développement végétatif des plantes ainsi qu'à l'estimation de la valeur fourragère :
- Nombre totale de feuilles (Nbr.Fll).
- Diamètre au collet (D.Coll) : Le diamètre au collet en (mm), mesuré à l'aide d'un pied à coulisse (figure 1).
- Longueur finale de la tige (Lg.Tige) : Mesurée en utilisant une règle graduée du collet à l'insertion du méristème apicale (figure 2).
- Poids de la matière sèche (PS) : Ce paramètre est nettement plus fiable et plus simple. La biomasse totale (en mg de matière sèche) est séchée dans l'étuve à 87°C puis pesée 48 heures plus tard.

Ces mêmes critères ont été aussi utilisés par d'autres auteurs dans l'estimation de la croissance de la biomasse chez certaines espèces d'*Acacia* cultivées sous stress salin [29, 30].

**Tableau 2:** Le tableau montre les paramètres mesurés dans l'évaluation de l'effet de la salinité au stade végétatif.

| Code des caractères | Signification |
|---|---|
| Lg.Tige | Longueur de la tige |
| D.Coll | Diamètre au collet |
| Nbr.Fll | Nombre totale de feuilles |
| PS | Biomasse totale (en mg de matière sèche) |

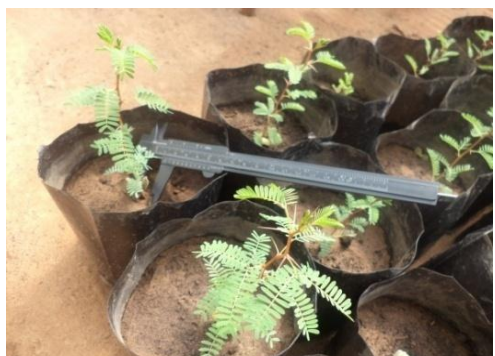
**Figure 1:** Mesure du diamètre au collet à l'aide d'un pied à coulisse.

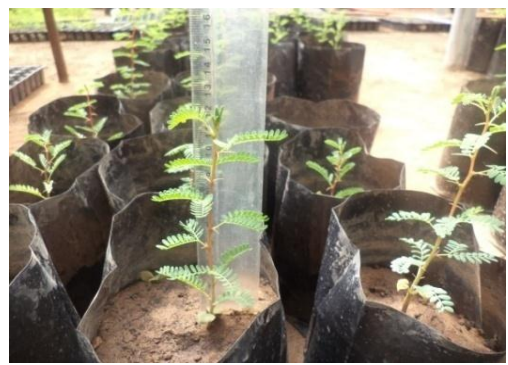
**Figure 2:** Mesure de la longueur de la tige à l'aide d'une règle graduée.

## 2.4. Analyses statistiques

Les six espèces ont été traitées selon un dispositif complètement randomisé, à raison de 10 plantes par population et par traitement. Les données relatives aux pourcentages de réduction des différents paramètres de croissance ont été transformées en arcsin racine carrée avant d'être soumises à l'analyse de variance à deux critères de classification (espèce et [NaCl]). La comparaison des moyennes entre les différentes espèces, pour chaque traitement, a été réalisée par le test de Newman et Keuls. Pour chaque concentration, les espèces dont les moyennes ne sont pas significativement différentes ont été regroupées dans une même ellipse sur les graphiques. Les traitements des données ont été réalisés par le logiciel Statistica (Version 6) [31].

# 3. RESULTATS

Le tableau 3 résume l'analyse de variance de l'effet espèce et de l'effet NaCl ainsi que leur interaction. Pour l'ensemble des caractères, le teste ANOVA montre un effet très hautement significatif entres les espèces et entre les différents traitements de sel. Ces différences sont plus marquées dans le cas des caractères se rapportant à la croissance en longueur des plantules. De la même manière, l'interaction (NaCl * Espèce) révèle aussi un effet très hautement significatif pour les caractères longueur de la tige et le nombre de feuilles et un effet significatif si on considère le poids total de la matière sèche tandis que pour le diamètre du collet l'interaction n'est pas significative. En conséquence, pour ce dernier critère, la salinité agit sur les différentes espèces de la même manière, quel que soit la concentration. Pour le reste des paramètres étudiés, l'effet de NaCl diffère d'une espèce à l'autre.

**Tableau 3:** Le tableau montre l'analyse de la variance à deux critères de classification abordée sur les différents caractères végétatifs mesurés chez les espèces d'*Acacia* étudiées.

| Caractères | $CM_{Espèce}$ | $CM_{Sel}$ | $CM_{Interaction}$ | $F_{Espèce}$ | $F_{Sel}$ | $F_{Interaction}$ |
|---|---|---|---|---|---|---|
| Lg.Tige | 11439,2 | 20596,0 | 555,2 | 121,697*** | 219,111*** | 5,907*** |
| D.Coll | 6051,7 | 2732,7 | 114,2 | 19,932*** | 9,000*** | 0,376 **NS** |
| Nbr.Fll | 3425,2 | 11918,6 | 317,0 | 32,291*** | 112,363*** | 2,988*** |
| PS | 6379,6 | 10101,6 | 279,3 | 42,478*** | 67,261*** | 1,860* |

**CM** : Carré moyen ; **NS** : test non significatif ; **\*** : test significatif ; **\*\*\*** : test très hautement significatif

## 3.1. Etude des caractères pris séparément pour l'ensemble des espèces étudiées
### 3.1.1. Longueur de la tige (Lg.Tige)

L'analyse de variance pour la longueur de la tige révèle un effet très hautement significatif des deux facteurs, salinité et espèce ainsi que leur interaction. Le classement des différentes espèces par le test de Newman et Kheuls selon leur tolérance à la salinité, montre une régression de la croissance en hauteur de la tige des plantules chez les différentes espèces étudiées (figure 3). Cependant, les réductions les plus importantes ont été notées pour *A. cyanophylla*, surtout dans le cas des fortes concentrations. Elle enregistre jusqu'à 92% pour une concentration de 400 mM. Par ailleurs, *A. horrida* semble la moins perturbée par le sel, du moins pour ce caractères, et affiche les valeurs les plus faibles. Les pourcentages de réduction varient dans ce cas entre 10 % et 47 %, respectivement pour les concentrations 100 et 400 mM. Le reste des espèces réagissent avec modération et occupent une situation relativement intermédiaire entre ces deux dernières espèces.

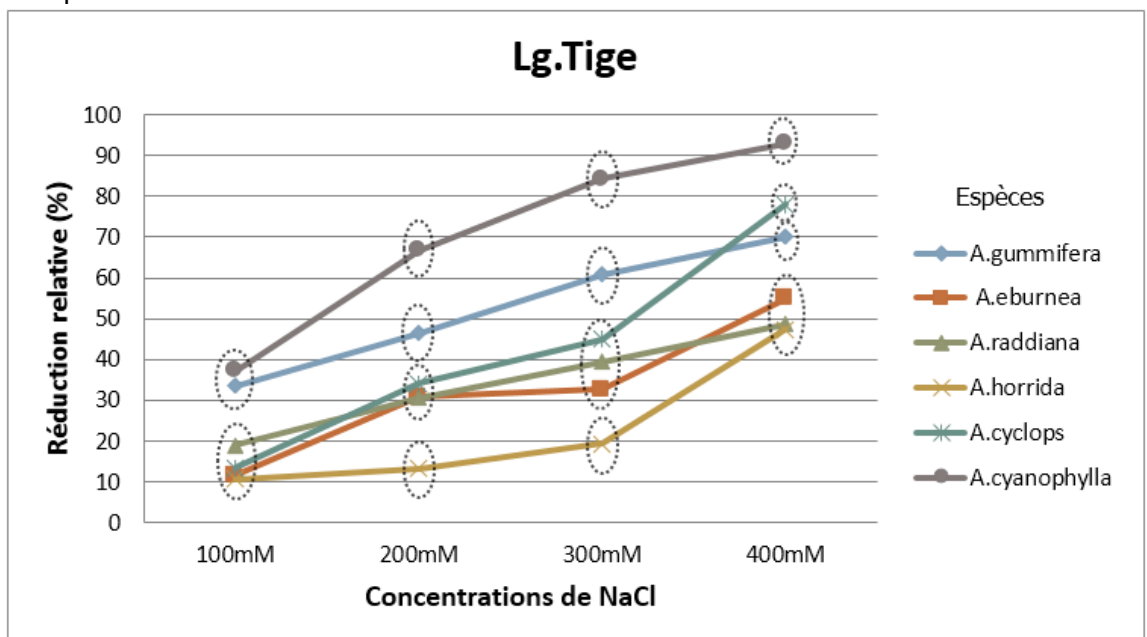

**Figure 3:** Représentation des moyennes calculées pour chaque espèce étudiées par traitement de NaCl pour la longueur de la tige. *(Les moyennes des espèces groupées dans la même ellipse ne sont pas significativement différentes selon le test de Newman et Keuls à 5%).*

### 3.1.2. Diamètre du collet (D.Coll)

L'augmentation de la concentration en sel dans la solution d'eau d'irrigation, diminue de façon significative le taux de croissance relative du diamètre du collet chez toutes les espèces considérées (figure 4). Pour ce critère on note une réduction plus faible par rapport aux autres caractères étudiés. À des concentrations allant de 200 à 400 mM, la réduction du diamètre du collet chez *A. gummifera* est significativement plus importante que celles des autres espèces. En effet, à 400 mM on note une régression qui peut atteindre 52% chez cette espèce comparativement à *A. horrida* chez laquelle la régression n'a été que de 14%. Les espèces *A. horrida*, *A. raddiana* et *A. cyanophylla* forment un groupe homogène et réagissent de la même manière au stress salin avec les réductions les plus faibles.

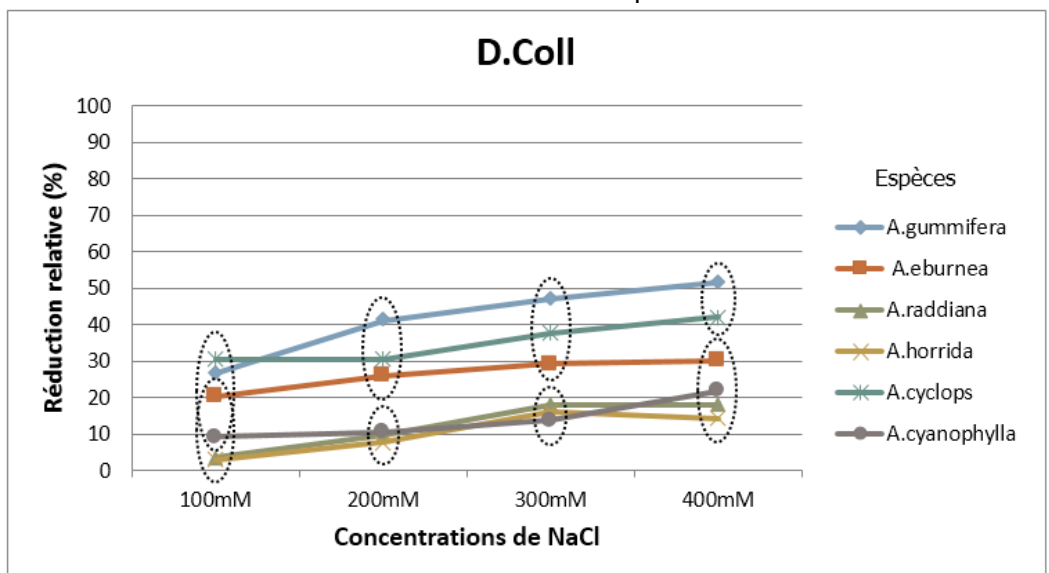

**Figure 4:** Représentation des moyennes calculées pour chaque espèce étudiées par traitement de NaCl pour le diamètre du collet. *(Les moyennes des espèces groupées dans la même ellipse ne sont pas significativement différentes selon le test de Newman et Keuls à 5%).*

### 3.1.3. Nombre de feuilles (Nbr.Fll)

Pour ce caractère, on note une classification moins nette des différentes espèces surtout au niveau des concentrations modérées en sel (figure 5). L'espèce *A. cyanophylla* semble encore une fois la plus affectée par le sel et montre une réduction de sa production foliaire atteignant les 86% à la concentration 400 mM de NaCl. En générale cette espèce exhibe la plus grande sensibilité à la salinité à ce stade de développement. D'un autre côté, *A. raddiana* a présenté une faible réduction du nombre de feuilles par plante à la concentration 300 mM de NaCl. On note, dans ce cas, une réduction de 33% chez l'espèce la plus tolérante contre 63% chez *A. cyanophylla* pour atteindre à la concentration de 400 mM une réduction de 48% chez *A. raddiana* et 86% chez l'espèce la plus sensible.

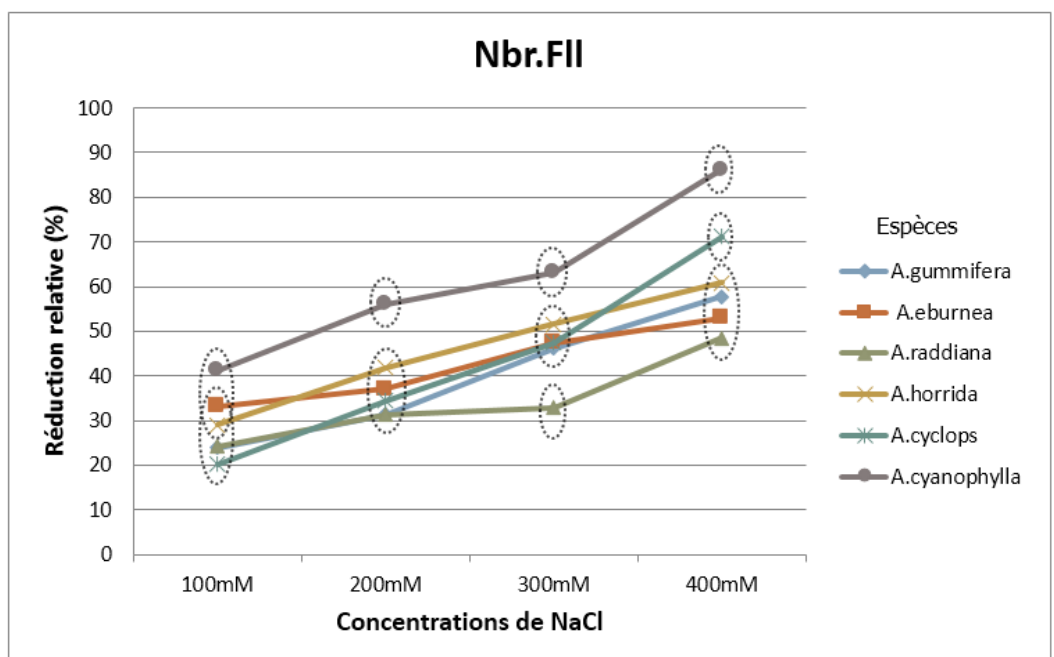

**Figure 5:** Représentation des moyennes calculées pour l'ensemble des espèces étudiées par traitement en NaCl pour le nombre de feuilles. *(Les moyennes des espèces groupées dans la même ellipse ne sont pas significativement différentes selon le test de Newman et Keuls à 5%).*

### 3.1.4. Biomasse totale (PS)

L'augmentation de la concentration de NaCl a un effet significatif sur la biomasse sèche de toutes les parties de la plante (feuilles, tiges et racines) des espèces testées (figure 6). La plus grande variabilité a été observée pour les deux concentrations 300 et 400 mM de NaCl. Comme pour les caractères nombre de feuilles et longueur de la tige, *A. cyanophylla* se distingue, encore une fois, des autres espèces en affichant pour ce critère, toutes concentrations confondues, les réductions les plus élevées. Par ailleurs, *A. horrida* et *A. raddiana* semblent être, en générale, moins affectées pour ce caractère, elles ont montrée de ce fait les plus faibles régressions de la matière sèche totale élaborée, variant en moyenne entre 13 et 34% contre 40 et 77% pour *A cyanophylla*, la plus sensible. Les autres espèces occupent une situation intermédiaire et montrent une grande variabilité au niveau des doses élevées en NaCl (400 mM) par rapport aux concentrations faibles.

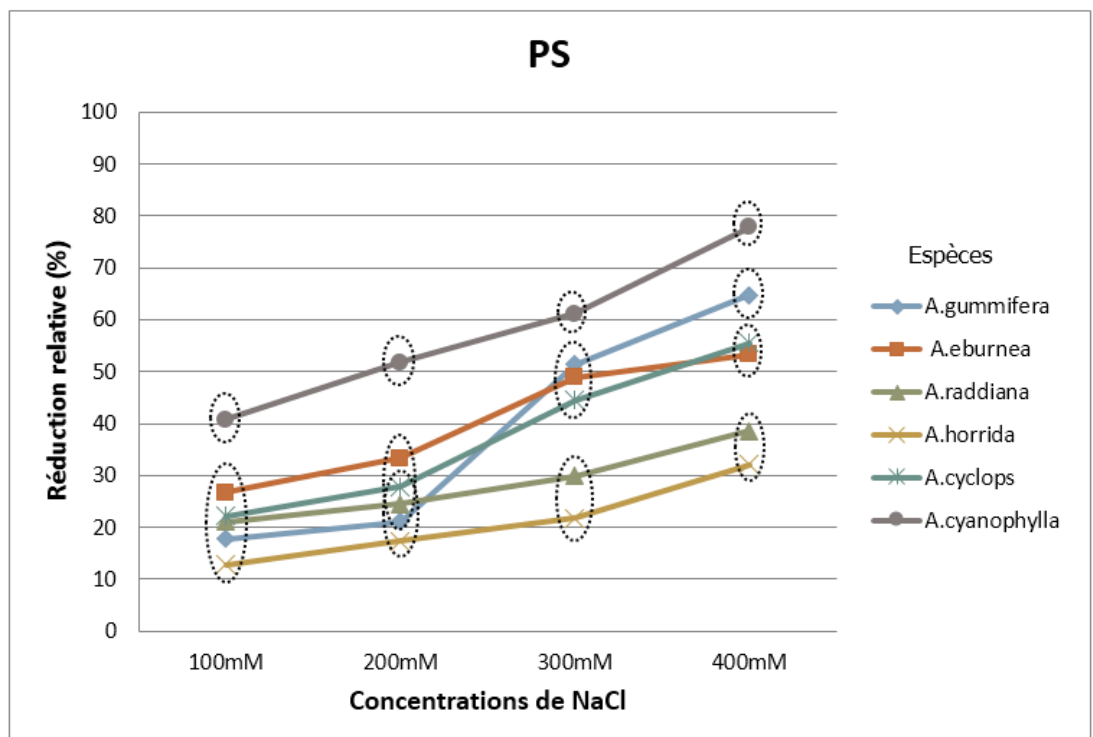

**Figure 6:** Représentation des moyennes calculées pour l'ensemble des espèces étudiées par traitement en NaCl pour la biomasse totale (PS). *(Les moyennes des espèces groupées dans la même ellipse ne sont pas significativement différentes selon le test de Newman et Keuls à 5%).*

## 4. DISCUSSIONS

Les résultats présentés dans cette partie, montrent que la salinité réduit en générale la croissance des plantules chez l'ensemble des espèces étudiées. Néanmoins, une grande variabilité entre les espèces a été révélée à ce stade. Les interactions très hautement significatives entre les deux effets (espèce*salinité), observées dans notre cas, montre la possibilité d'une sélection essentiellement sur la base des caractères : Longueur de la tige, nombre de feuille et la biomasse totale. Toutefois, on a constaté qu'une espèce performante pour un caractère donné n'est pas forcément la meilleure pour un autre critère. C'est le cas d'*A. cyanophylla* dont le nombre de feuilles, la biomasse totale et la longueur de la tige paraissaient relativement affectée par le sel, mais dans le cas du caractère relatif au diamètre du collet elle était classée parmi les espèces les plus tolérantes.

Nos résultats sont en concordance avec les travaux de Nguyen et al. [32] dans lesquels ils ont révélés que les deux espèces *Acacia auriculiformis* et *Acacia mangium* ont réagi également par une réduction de la croissance de la partie aérienne en réponse au stress salin. Cet effet est fréquent chez les glycophytes [33], où la diminution de la croissance de l'appareil végétatif observée peut être expliquée par une augmentation de la pression osmotique provoquée par NaCl, ce qui bloque l'absorption de l'eau par les racines. Les plantes s'adaptent ainsi au stress salin par la réduction de leur croissance afin d'éviter les dommages causés par le sel [34, 35]. Les effets de la salinité sur la croissance des plantules cultivés en conditions semi contrôlées, dépendent de plusieurs facteurs. Ils varient selon la teneur de NaCl appliquée, l'espèce, la provenance, le stade végétatif et la partie de la plante [2]. Les effets de la salinité se manifestent principalement par un ralentissement de la croissance de l'appareil végétatif.

D'autre part, il est important de signaler que toutes les espèces d'*Acacia* considérées dans ce travail ont survécu, même à 400 mM de NaCl, alors que selon Ghulam et al. [36], ce niveau de salinité a été nuisible pour *A. nilotica* tandis que *A. ampliceps* tolère cette concentration. La comparaison de la tolérance à la salinité chez cinq espèces d'*Acacia* : *A. ampliceps, A. salicina, A. ligulata, A. holosericea* et *A. mangium* a révélé que *A. ampliceps* était la plus tolérante et a survécu même à 428 mM en NaCl, concentration à laquelle toutes les autres espèces ont été sévèrement affectées [37].

Pour le paramètre matière sèche, les réductions les plus faibles sont enregistrées chez *A. horrida* et *A. raddiana*. Le pourcentage de réduction de la matière sèche, généralement considéré comme indice de sensibilité des plantes vis-à-vis du stress salin, montre que la concentration 400 mM NaCl est insuffisante pour engendrer une réduction relative de 50 % par rapport aux témoins, seuil très utilisé pour le classement de la tolérance des plantes [21]. En outre, même à une concentration plus faible (300 mM) les autres espèces manifestent des réductions plus marquées.

Les feuilles sont les parties les plus sensibles de la plante au stress salin. Cependant, chez l'ensemble des espèces on assiste à une réduction significative du nombre de feuilles par rapport au témoin surtout à partir de 300 mM de NaCl. Des résultats similaires ont été rapportés sur d'autres espèces par [38-41]. D'autre part, au cours de l'expérimentation on a constaté que la croissance foliaire est également très affectée par l'augmentation du stress salin quelle que soit l'espèce. L'expansion des feuilles est considérablement inhibée par le sel, les nouvelles feuilles se développent lentement et le vieillissement des feuilles âgées s'accélère. D'ailleurs, quand la surface foliaire est réduite par la salinité, la production de carbohydrates devient insuffisante pour supporter la croissance et le rendement [42].

La réduction de la croissance, dans les conditions d'un stress salin est attribuée à plusieurs facteurs, parmi lesquelles l'accumulation des ions, aussi bien en $Na^+$ qu'en $Cl^-$ à des teneurs élevées dans les tissus foliaires qui est la cause principale des contraintes ioniques au niveau des tissus de la plante [43]. Selon ces auteurs, le stress salin cause un déséquilibre nutritionnel qui en résulte l'inhibition de l'absorption des éléments nutritifs essentiels comme le $Ca^{2+}$, $K^+$, $Mg^{2+}$, $NO_3^-$ par les phénomènes de compétition minérale de fixation apoplasmique [44]. En outre, il est établi qu'un supplément de $Ca^{2+}$ dans le milieu de culture améliore les conditions de croissance sous stress salin [45]. Le dérèglement de l'absorption du calcium inhibe également l'établissement de la nodulation et la fixation d'azote chez les légumineuses [46]. Il parait que cet ion est impliqué dans le processus de la reconnaissance *Rhizobium*-poil absorbant [47].

Par ailleurs, la diminution de la croissance des parties aérienne peut aussi être expliquer par des perturbations des taux des régulateurs de croissance dans les tissus, particulièrement l'acide abscissique et les cytokinines induites par le sel [48], mais aussi à une diminution de la capacité photosynthétique provoqué par la diminution de la conductance stomatique de $CO_2$ sous la contrainte saline. Dans tous les cas, cette réduction de la croissance des différentes parties aériennes est considérée comme une stratégie adaptative nécessaire à la survie des plantes exposées à la salinité [49]. Ceci permet à la plante d'emmagasiner de l'énergie nécessaire pour faire face au stress afin de réduire les dommages irréversibles occasionnés, quand le seuil de la concentration létale est atteint [39].

De plus, il a été constaté que la tolérance au stade germination, dans les conditions de nos expériences, ne reflète pas dans tous les cas celle au stade végétatif. En effet, l'espèce *A. horrida,* classée parmi les plus sensibles au sel durant la germination, manifeste une tolérance importante vis-à-vis de NaCl au stade plantule. Par contre l'espèce *A. raddiana,* reconnue par sa tolérance à la salinité, très marquée, au stade germination, conserve globalement sa performance aux stades avancés. Cependant, la germination sous contrainte saline n'est pas suffisante pour identifier des espèces tolérantes au sel [26, 50]. Dans ce contexte, de nombreux auteurs ont montré que la réponse à la salinité variait selon le stade de développement de la plante [51-53]. Toutefois, la germination et les premiers stades de la croissance seraient les phases les plus sensibles [54].

## 5. CONCLUSION

Le stress salin exerce chez les Six espèces d'*Acacia* un effet dépressif sur tous les paramètres de croissance étudiés. Toutefois, le taux de réduction diffère selon l'intensité de stress salin et le degré de sensibilité ou de tolérance de l'espèce. La croissance en hauteur, le nombre de feuilles et la biomasse sèche totale sont vraisemblablement les plus affectés. Cependant, toutes les espèces d'*Acacia* considérées dans ce travail ont survécu, même à 400 mM de NaCl et présentent différents degrés de tolérance à la salinité. Les espèces *A. horrida* et *A. raddiana* s'avèrent globalement les plus performantes. Cette variabilité génétique dévoilée par ces espèces dans les différentes conditions de stress salin, surtout sous des seuils élevés en NaCl allant jusqu'à 400 mM, constitue un atout intéressant qui peut être utilisé aussi bien dans le choix des espèces à retenir pour améliorer la tolérance à la salinité que dans les programmes de valorisation et de réhabilitation des sols salés.

Par ailleurs, ces recherches doivent être poursuivies par des expériences à des stades végétatifs plus avancés afin de confirmer les résultats constatés aux stades germination et juvénile. Toutefois, le dosage et l'identification des osmorégulateurs tels que la proline, pourrait mieux éclaircir les mécanismes d'ajustement osmotique nécessaires à ces plantes pour s'adapter au stress salin. Cela pourrait expliquer, par la même, leur tendance à l'halophilie, observée au cours des deux stades étudiés.